%% file: main.tex
\documentclass[runningheads]{llncs}

\usepackage[disable]{todonotes}
\usepackage{graphicx}
\usepackage[utf8]{inputenc}
\usepackage{xcolor}
\usepackage{amsmath,amssymb}
\usepackage[inline]{enumitem}

\usepackage{caption}
\usepackage{subcaption}
\usepackage{multirow}

\newcommand\hmm[1]{\ifnum\ifhmode\spacefactor\else2000\fi>1000 \uppercase{#1}\else#1\fi}

\newcommand{\hide}[1]{}
\newcommand{\reminder}[1]{{\todo{#1}}}

\begin{document}
\title{Pattern-based Visualization of Knowledge Graphs}

\author{Luigi Asprino\inst{1} \and
Christian Colonna\inst{1} \and
Misael Mongiovì\inst{2} \and Margherita Porena \inst{2} \and Valentina Presutti \inst{1,2}}
\authorrunning{L. Asprino et al.}

\institute{University of Bologna, Bologna, Italy \and
STLab, ISTC-CNR, Rome, Italy \\
\email{\{luigi.asprino,christian.colonna2,valentina.presutti\}@unibo.it} \\
\email{\{misael.mongiovi,margherita.porena\}@istc.cnr.it}}

\maketitle              
\begin{abstract}
We present a novel approach to knowledge graph visualization based on ontology design patterns. 
This approach relies on OPLa (Ontology Pattern Language) annotations and on a catalogue of visual frames, which are associated with foundational ontology design patterns. 
We demonstrate that this approach  significantly reduces the cognitive load required to users for visualizing and interpreting  a knowledge graph and guides the user in exploring it through meaningful thematic paths provided by ontology patterns.




\keywords{Knowledge Graphs \and Information Visualization  \and Ontology Design Patterns \and Exploratory Search.}
\end{abstract}

\section{Introduction}
\label{sec:intro}
\input{src/intro}

\section{Related Work}
\label{sec:related}
\input{src/related}

\section{Approach}
\label{sec:approach}
\input{src/approach}

\section{Implementation}
\label{sec:implementation}
\input{src/implementation}

\section{User Study}
\label{sec:evaluation}
\input{src/evaluation}


\section{Conclusions and Future Work}
\label{sec:conclusions}
\input{src/conclusions}

\bibliographystyle{splncs04}
\bibliography{biblio}

\end{document}

%% file: src/intro.tex
Exploring and understanding large knowledge graphs is a recurrent and tricky task in most knowledge graph and ontology projects. Its difficulty is amplified by the magnitude and the heterogeneity of the analyzed KG~\cite{DBLP:journals/fgcs/DesimoniP20}. Visualization methods are the main means to support KG understanding, but existing visualization tools mainly reflect the syntactic structure of KGs rather than their conceptual semantics, thus limiting usability and clarity~\cite{DBLP:journals/ker/DudasLSP18}. An ideal KG visualization tool should facilitate the answering of questions such as: what are the conceptual components of a KG? What are the key elements of such components? How these components are instantiated? Which altogether help answering the more general question of what a KG \textit{talks about}. We think that to address this problem there is a need for a change of perspective on how to visualize the content of a KG, and that this perspective can be supported by ontology design patterns. Ideally, an intuitive (displayed) overview of a large knowledge graph should fit the dimension of a page, which can be achieved if KG data are grouped according to conceptual components, identified by ODPs.
%

In this paper, we propose a novel approach to KG visualization which relies on the notion of Ontology Design Pattern.
Ontology Design Patterns (ODPs)~\cite{gangemi2009ontology} are modeling solutions that address recurrent ontology design problems. 
ODPs are the basic building blocks of ontologies, and, consequently, they  shape  the (ontology-based) KGs.
In this vision,  ODPs embody the key conceptual components that a KG instantiates.
Following this intuition, we devised an approach that promotes ontology design patterns as first-class citizens for accessing and navigating KGs.
The approach relies on the following hypotheses.
\begin{enumerate*}[label=\textbf{(H\arabic*)}]
    
    \item \label{summary} The collection of ODPs provides a very concise summary of the overall content of a KG.
    
    \item \label{view} Each ODP provides a meaningful view of the content of a KG.
    
    \item \label{path} ODPs define thematic paths that guide the exploration and interaction with the KG.
    
    \item \label{visualization} Each ODP can be associated with  a visual frame that is meant to define an intuitive standard visualization for the instances of a pattern thus easing the interpretation of the information.
    
\end{enumerate*}

To assess the feasibility and benefits of our approach we conducted a pilot study.
In this study, we developed a tool, called ODPReactor, which provides us with a proof-of-concept of the pattern-based approach to KG visualization.
ODPReactor leverages the Ontology Pattern Language annotations (OPLa) for recognizing instances of patterns within a KG.
Once pattern instances are recognized, ODPReactor summarizes the content of the KG in a single concise visualization showing what are the patterns composing the KG,  how they are instantiated and what are the key concepts of each pattern (cf. \ref{summary}).
ODPReactor allows the users to select a pattern for accessing and navigating the KG (cf. \ref{view} and \ref{path}).  
ODPReactor also defines a set of visual frames that are used to display intuitively the data in a pattern occurrence (cf. \ref{visualization}).

As a case study, we deployed ODPReactor for accessing (a portion of) ArCo, the Italian Cultural Heritage KG~\cite{carriero2019arco}.
ArCo consists of a network of seven ontologies and 169 million triples about 820 thousand cultural entities.
ArCo's ontologies reuse ODPs from online repositories (e.g. ODP portal) whose occurrences are annotated using OPLa.
We selected the ODPs to ensure a balance between patterns involving ``physical''   and ``abstract'' concepts (for example, the visualization of the geo-localization of a cultural property involves physical objects only, while a collection of measurements of an object involves both physical and abstract concepts) and then, we associated  each pattern with a cognitive-grounded visualization.
Finally, to validate our hypotheses, we conducted a user study that involved 11 participants. The results show the validity of the ODP-based visualization approach in terms of usability and rapidity in accessing the information.

The remainder of this paper is structured as follows. After a discussion of related work (Sect. \ref{sec:related}), we present the proposed ODP-based visualization approach (Sect. \ref{sec:approach}) and discuss its implementation (Sect. \ref{sec:implementation}). Eventually we report the results of our user study (Sect. \ref{sec:evaluation}) and conclude the paper.

%% file: src/related.tex




We place our research in the area of Knowledge Graph visualization.
To the best of our knowledge, the only tool that adopts patterns as guidance for the exploratory search is \emph{Aemoo}~\cite{nuzzolese2017aemoo}.
Aemoo is a Linked Data exploration tool that uses Encyclopedic Knowledge Patterns as relevance criteria for selecting, organizing, and visualizing knowledge.
Although our work draws inspiration from Aemoo, our approach aims at broadening Aemoo's scope (which is limited to encyclopedic patterns) by developing a general method for interacting with Knowledge Graphs shaped according to any Ontology Design Pattern.

Plenty of approaches have been proposed for visualizing KGs~\cite{desimoni2020empirical}. 
Most of them fall into two categories: \emph{graph-based visualization} and \emph{template-based visualization}~\cite{desimoni2020empirical}. 
Tools in the former category use network layouts to visualize RDF data as they mirror the underlying RDF structure. 
Tools belonging to the latter category rely on the design of predefined HTML templates which are populated with resources resulting from the execution of predefined SPARQL queries.


\noindent
\textit{Graph-based visualization tools.}
\emph{H-BOLD}~\cite{po2018high} is a graph-based visualization tool whose architecture is divided in three different levels: the Cluster View, the Schema View, and the Instance View.
This approach is similar to ours, especially in the summarization provided in the Cluster View.\todo{Christian Controlla che non abbia scritto cavolate. [CC:] usano Louvain community detection algorithm. Mi pare vada bene dire "is guided by statistical correlation" }
However, the summarization is guided by statistical correlations instead of conceptual analysis of the KG.
\textit{KC-Viz}~\cite{motta2011novel} was among the firsts ontology\reminder{occhio questi due sono per ontology visualization, ho cambiato.} visualization tools using the notion of \textit{key concepts}~ \cite{peroni2008identifying} (which are the statistically relevant classes for the KG at hand).
Other tools (e.g. \emph{RDFDIGEST+}~\cite{troullinou2018rdfdigest+}) followed the KC-Viz approach.
We adopt a similar strategy for selecting the key concepts of a KG and relating them to ODPs.


\noindent
\textit{Template-based visualization tools.}
Exhibit~\cite{huynh2007exhibit} is one of the first example of usage of templates (called Lens Templates) for presenting KGs.
On the same line of research we can find other works based on template based visualization, such as \emph{LodView}\footnote{\url{https://lodview.it/}} (the tool adopted for accessing the ArCo KG) or others~\cite{arndt2019jekyll,wang2018template,luggen2015uduvudu,thellmann2015linkdaviz,auer2010less}.
A special case is the tool proposed by McBrien et al.~\cite{mcbrien2019conceptual} which includes a mechanism for recommending the most appropriate visualization for a given SPARQL query.
Our approach leverages these experiences, meaning that, the visualization strategies implemented by  ODPReactor could be seen as templates for visualizing a KG.
A limit of the existing template-based approaches is that they consider as meaningful boundary for describing an entity its  neighborhood only (which is typically the set of triples involving a resource), and all the entities are described with the same boundary.
However, while this is an effective strategy for ``simple'' KGs, it fails short with sophisticated KGs in which the meaningful boundary of each entity depends on the type of the entity and the pattern the entity is involved in.
ODPReactor goes beyond these approaches since it associates each entity of the KG with a meaningful boundary that depends on the ODPs the entity is involved in (implying that each entity might have a different boundary).

%% file: src/approach.tex
In this section we briefly review the main concepts we build upon (ODP, OPLa and key-concept classes), then we describe our framework and an example of visualization based on the ODP Part Of. 

\subsection{Background}\label{sec:background}

\textbf{Design patterns:} Ontology Design Patterns (ODPs)\footnote{\url{http://ontologydesignpatterns.org}} \cite{gangemi2009ontology} are modeling solutions to recurrent ontology design problems. ODPs can be divided into multiple categories: \emph{logical}, \emph{architectural}, \emph{content}, \emph{presentation} \cite{presutti2008content}.
In our case study, we used three ontology design patterns (implemented in the context of ArCo project \cite{carriero2019pattern}). The first, \emph{Time-Indexed Typed Location}, represents the locations of an object (e.g. a cultural property) in specific time intervals and with specific purposes (e.g. storage, exposition). It specializes \emph{Time-Indexed Situation}, a pattern to represent situations that have an explicit time parameter. The second, \emph{Measurement Collection}, represents a collection of measures. It is a specialization of pattern \emph{Collection}, to represents collections and their members. Last, \emph{Cultural Property Component Of} represents a cultural property and its components and specializes \emph{Part Of}, a pattern to represent entities and their parts. 

\textbf{OPLa:} OPLa ontology is an ontology design pattern representation language \cite{hitzler2017towards}, which enables annotating ontologies to (1) indicate that an axiom or class belongs to a certain module or pattern, (2) 
specify how an axiom or class is related to the pattern,
and (3) represent relationships between two patterns or between a pattern and a module.
We extended OPLa to represent instances of patterns.
We added the property \texttt{opla:isPatternInstanceOf} with domain an individual of type \texttt{opla:PatternInstance} and range an individual of type \texttt{opla:Pattern}. To indicate that an entity belongs to a pattern instance. We defined the property \texttt{opla:belongsToPatternInstance} (and its inverse \texttt{opla:hasPatternInstanceMember}), whose domain and range are respectively \texttt{owl:Thing} and \texttt{opla:PatternInstance}.
Our methodology relies on the use of this annotations to identify and distinguish specific instances of patterns in data.


\textbf{Key Concepts:} Knowledge graph summarization is crucial in graph visualization and other applications where large graphs need to be handled
\cite{vcebiric2019summarizing}. 
In \cite{peroni2008identifying} the authors introduced \emph{key concepts} as elements which best summarize an ontology. 
Following this approach, we adopted the Degree centrality measure~\cite{vcebiric2019summarizing} (a measure often used for knowledge graph summarization~\cite{troullinou2018exploring}) for estimating the relevance of each class of a given ontology.

\subsection{Pattern-based visualization of Knowledge Graphs}

We propose a general framework based on three levels of visualizations. 
The top level, namely the \emph{ODP level}, is a view of the knowledge graph that shows a graph whose nodes represent either an ODP instantiated in the KG or a key concept of an ODP.
Edges of this graph are modeled according to the OPLa vocabulary and its extension discussed in Sect. \ref{sec:background}. 
The second level, called \emph{exploration level}, presents the occurrences of a specific ODP and allows users to filter the occurrences according to her information needs.
The third one, the \emph{visualization level}, is dedicated to the visualization of a single occurrence of an ODP or an entity of the KG. 
First, we intuitively describe the aim and the characteristic of these levels, and, then we formally define our visualization method.

We remark that the design of all the layers followed Gestalt psychology principles \cite{kohler1967gestalt}
and studies on visual language \cite{horn1998visual}. 
This principles were particularly relevant for the visualization level.
For example, Common Region Gestalt principle states that when objects are enclosed in the same region we perceive them as being grouped together. Following this idea, we enclosed all members of a collection within a line in designing the visual frame for pattern \emph{Collection}. 

\begin{figure}[t!]
  \centering
  \begin{subfigure}[b]{0.30\linewidth}
    \includegraphics[width=\linewidth]{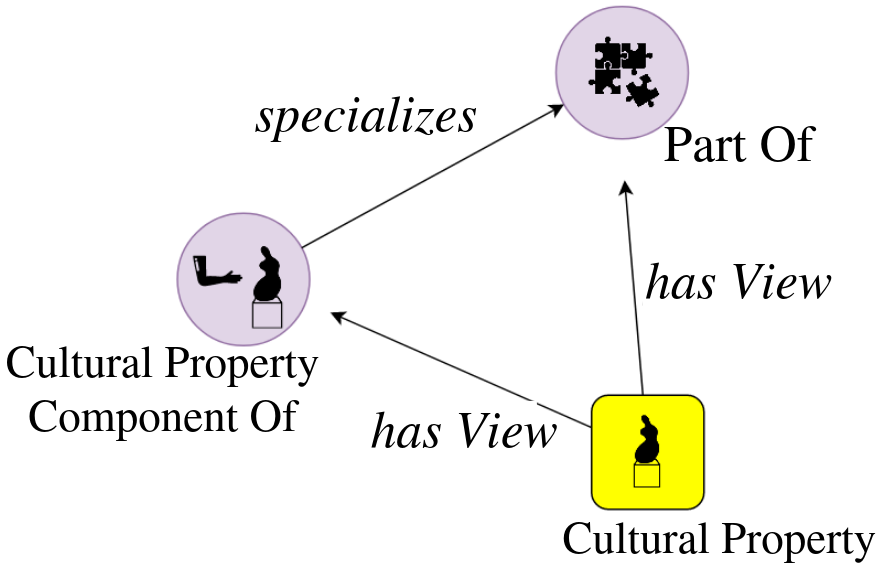}
     \caption{ODP level: graph of ODPs and key concepts}\label{fig:odp_level}
  \end{subfigure}
  ~
  \begin{subfigure}[b]{0.42\linewidth}
    \includegraphics[width=\linewidth]{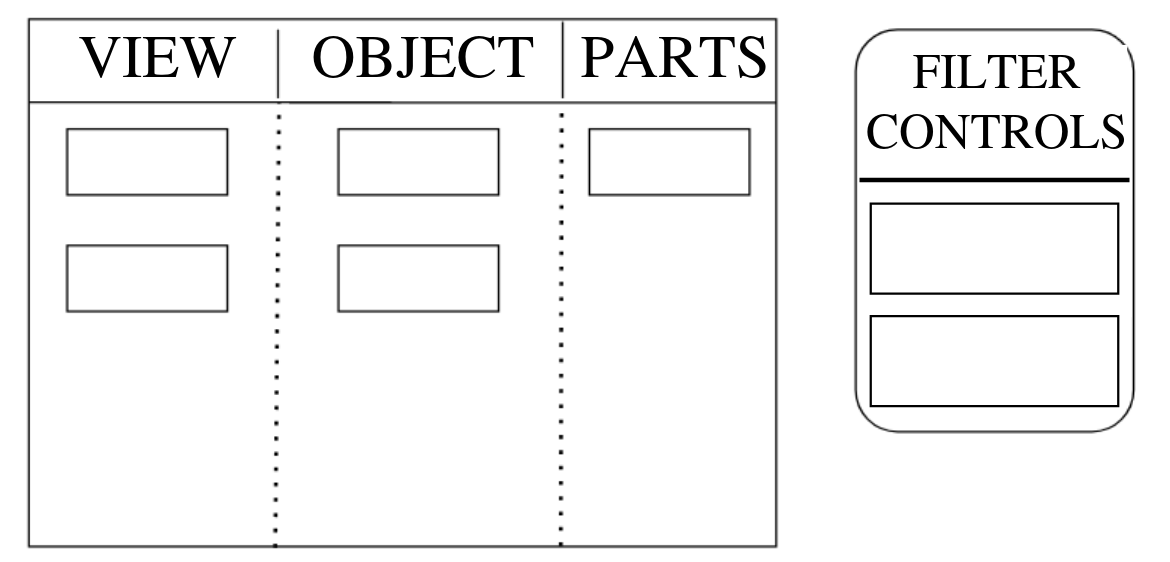}
    \caption{Exploration level: pattern instances and filter controls}\label{fig:exp_level}
  \end{subfigure}
  ~
  \begin{subfigure}[b]{0.23\linewidth}
    \includegraphics[width=\linewidth]{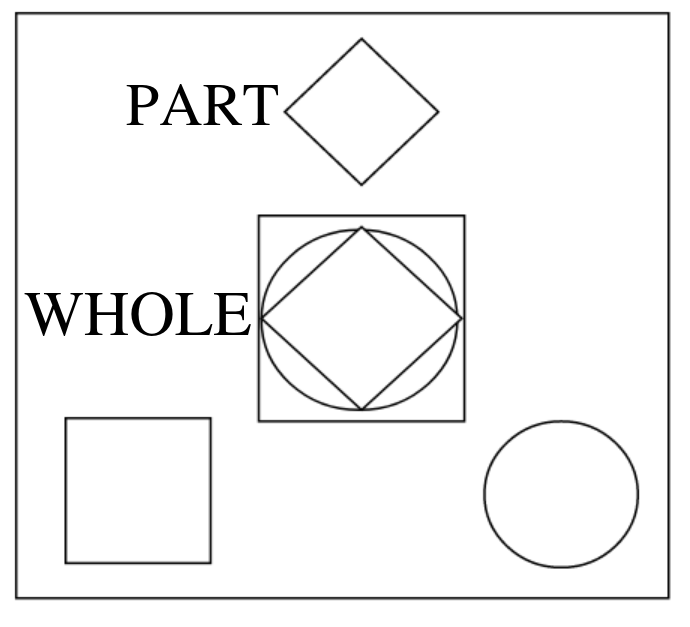}
    \caption{Visualization level
    }\label{fig:vis_level}
  \end{subfigure}
  \caption{Overview of the system on three levels}
  \label{fig:coffee3}
\end{figure}

\noindent
\textbf{ODP level.} This is the highest level of abstraction and summarization of a knowledge graph.
This visualization provides (in the dimension of a page) a concise overview (cf. \textbf{H1}) of the most relevant concepts instantiated in a  KG.
Specifically, the system shows a graph representation of the most relevant key concepts (i.e. classes) of a KG and a number of \emph{views} related to such concept.
Each view corresponds to an ODP and identifies a perspective for exploring a KG.
Views and key concepts are represented by nodes, while edges explain the relations among them. 
Figure \ref{fig:odp_level} depicts an example of the graph visualized at this level. 
The graph contains two views (Part Of and its specialization Cultural Property Component Of) and a key concept (Cultural Property). Edges are of two types: those describing the relation between ODPs (e.g. specializes, has component) and those relating a key concept to the ODP it appears in (has View).
It is worth noticing that both relations are retrieved from OPLa annotations, the edge \textit{has view} symbolizes \texttt{opla:isNativeTo} property, whereas specializes edges stand for  \texttt{opla:specializationOfPattern}.
ODPReactor also provides filters (not shown in Figure) to reduce the number of nodes in the graph visualization, including an importance threshold for classes to be visualized as key concepts. 
the size of the nodes is proportional to the logarithm of the
number of occurrences of the pattern.
This provides the user with an insight of how the various components of the KG are instantiated.
Interestingly, this view gives a concise overview of the overall knowledge graph 
From this view the user can rapidly perceive a summary of the information in the KG: what its main conceptual components are, what the key elements of such components are and how these components are instantiated. 


\noindent
\textbf{Exploration level.} The Exploration level is the second level of visualization. 
This level shows the instances of ODPs and key concepts.
The user can access to this level by clicking either on an ODP or a key concept showed at the ODP level.
In both cases instances are presented in a tabular form. A schematic representation of this level is shown in Figure \ref{fig:exp_level}. As for the instances of the key concepts, the table shows the list of individuals that can be filtered by the user.
The user can access to the detailed view of a single individual by clicking on a row in the table.

As for ODP instances, each column of the table is associated with an ODP dimension (e.g. for Time Indexed Typed Location dimensions are start time, end time, location type, and coordinates), while each row represents an instance. 
For every dimension of an ODP, we designed a set of \emph{semantic filters} that allow users to define the criteria for filtering instances. 
In agreement with the \emph{open-world assumption}, by default, the filters include instances that do not have a value for the filtering criteria.
However, we give the possibility to switch to the \textit{close-world assumption} and exclude resources for which missing data do not consent filters to be applied. 
Clearly, filters can be combined in order to restrict the set of instances according to multiple criteria defined on multiple dimensions.
Then, the user can access to the detailed view of a single pattern instance by clicking a row of the table.

It is noteworthy that the exploration and the filtering strategy implemented in the visualization depends on the ODPs which the KG is composed of.
Therefore, both the summary presented at ODP level and the interaction designed at visualization level rely on the ODPs instantiated in the KG thus complying with the hypotheses \textbf{H2} and \textbf{H3}.

\noindent
\textbf{Visualization level.} 
The visualization level provides the users with a detailed view of an instance of a pattern or a class of the KG.
Specifically, this visualization presents the pattern instances through the lens of a visual frame.
A visual frame is a standard visualization designed for a pattern (e.g. Geo-Localization of entities are depicted as markers of a map) which  presents the pattern instance by means of a set of intuitive graphical elements. Figure \ref{fig:vis_level} depicts a schematic example of visual frame for pattern Part Of.
It is worth noticing that, even if the visual frame is  specifically designed for an ODP, it can be reused on different KGs implementing the ODP.
Consequently, this strategy implicitly promotes  visual frames as standard visualization for a pattern (cf. \textbf{H4}).
The visual frames are also displayed in the visualization of the individuals of the KG.
In fact, the visualization level for an individual is designed for displaying all the visual frames which the individual is involved in and all the property-value pairs associated to the individual.
Interestingly, in doing so, the boundary of information presented together an individual of the KG is delimited by the boundary of the ODPs.


\noindent
\textbf{Formalization of the approach.} We now formally describe our approach. We define a RDF knowledge graph $\mathcal{G}$ as $\mathcal{G} \subseteq (\mathcal{U} \cup \mathcal{B}) \times \mathcal{U} \times (\mathcal{U} \cup \mathcal{B} \cup \mathcal{L})$  where $\mathcal{U}$, $\mathcal{B}$ and $\mathcal{L}$ are the set of URIs, the set of blank nodes and the set of literals, respectively. We assume that ODPs are annotated with OPLa ontology~\cite{hitzler2017towards}. Briefly, classes that belong to an ODP are annotated with a relation \texttt{opla:isNativeTo} connecting to the URI of the corresponding ODP pattern. We define $\mathcal{P}$ as the set of ODP patterns implemented in $\mathcal{G}$. We also define $implementation_{\mathcal{G}}(P) \subseteq \mathcal{G}$ as the subgraph (set of tuples) corresponding to the TBox implementation of the pattern in the knowledge graph, induced by the set of source nodes of relations \texttt{opla:isNativeTo} targeted in $P$. We also considers a special set of (\emph{key concept}) nodes $\mathcal{K}$, which are special classes that help the user identify and distinguish specific instances of patterns.

Given an ODP pattern $P \in \mathcal{P}$, we can identify a number of occurrences of such a pattern in the knowledge graph, where an occurrence is an ABox implementation of $P$, composed by individuals and relations that respect the pattern structure. The identification of such occurrences is dependent on the specific pattern and can be performed by suitable SPARQL queries. While a general way to identify ODP occurrences is worth of further investigation, in this work we assume such occurrences have been previously identified and annotated with the extension of OPLa discussed in Sect. \ref{sec:background}\footnote{We implemented the code to generate such annotations for the patterns we employed in our prototype}. We denote the set of occurrences of a pattern $P$ as $\mathcal{Q}_P$, where an element $Q \in \mathcal{Q}_P$ is a subgraph of $\mathcal{G}$ that concretely materializes $P$. Their nodes are denoted as $nodes(Q)$.

At the ODP level our tool shows a graph whose nodes represent patterns and key concepts ($\mathcal{P} \cup \mathcal{K}$) and edges show their relations. A Key-concept node $K\in \mathcal{K}$, which corresponds to a class in $\mathcal{G}$, is connected to a pattern node $P$ through a relation \texttt{hasView} if $K$ is a class in $implementation_\mathcal{G}(P)$. Patterns are also interconnected by OPLa \cite{hitzler2017towards} relations (e.g. \texttt{opla:specializationOfPattern} -- in the tool URIs are substituted with user-friendly descriptions).   

The exploration level requires a pattern-specific mapping function from data elements to visual elements and a pattern-specific set of filter controls. Since at this level data are visualized in a table, the mapping needs to associate nodes of the pattern (typically literals) to columns. Specifically, each pattern $P$ is associated to a set of $c$ columns and a function (implemented by specific code) $emap_P : \mathcal{Q}_P \rightarrow \mathcal{L}^{c} \cup \{\phi\}$ which associates a pattern instance $Q \in \mathcal{Q}_P$ to a list of $c$ literals to be shown in a row. The special symbol $\phi$ accounts for filters. Specifically $emap_P(Q)=\phi$ if $Q$ does not satisfy the filter constraints set by the user. The GUI for filters is pattern dependent and can be built by combining a set of filter controls, each one associated to a specific property. We implemented a set of filter control templates to account for different data types (dates, locations, numerical values, discrete values), which can be combined to construct the filter GUI of each pattern.

The visualization level consists of the visualization of a pattern instance in a specific page. Our approach requires a pattern-specific function $vmap_P : \mathcal{Q}_P \rightarrow \mathcal{VF}$ that associates each pattern instance $Q$ to its correspondent visualization frame. The correspondence between data values and visual elements is established by suitable knowledge-graph-independent SPARQL queries.

\begin{figure}[t]
     \centering
     \begin{subfigure}[b]{0.55\textwidth}
         \centering
         \includegraphics[width=\textwidth]{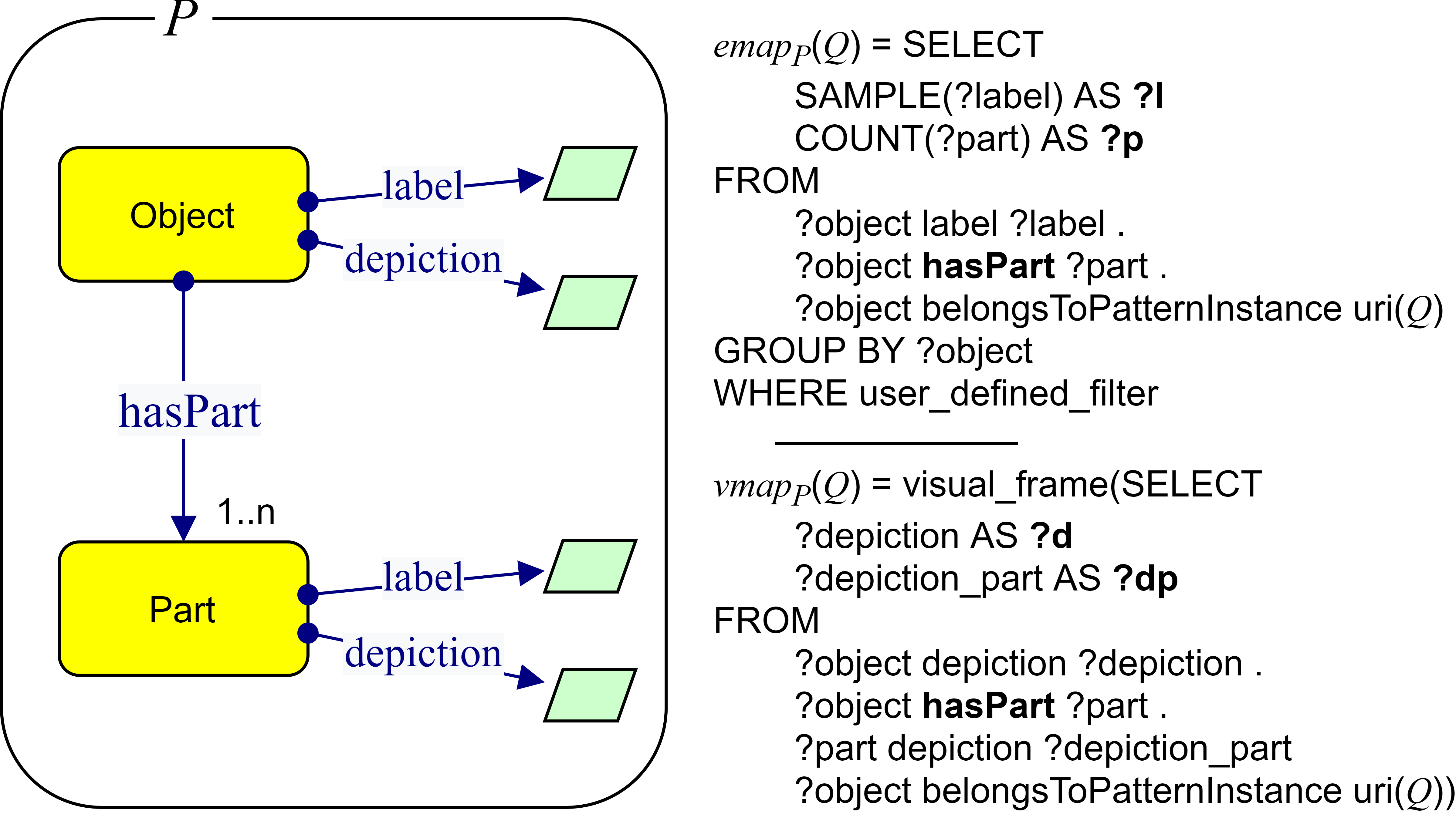}
         \caption{Pattern Part Of and associated mapping functions}
         \label{fig:partof_pattern}
     \end{subfigure}
     \hfill
     \begin{subfigure}[b]{0.44\textwidth}
         \centering
         \includegraphics[width=\textwidth]{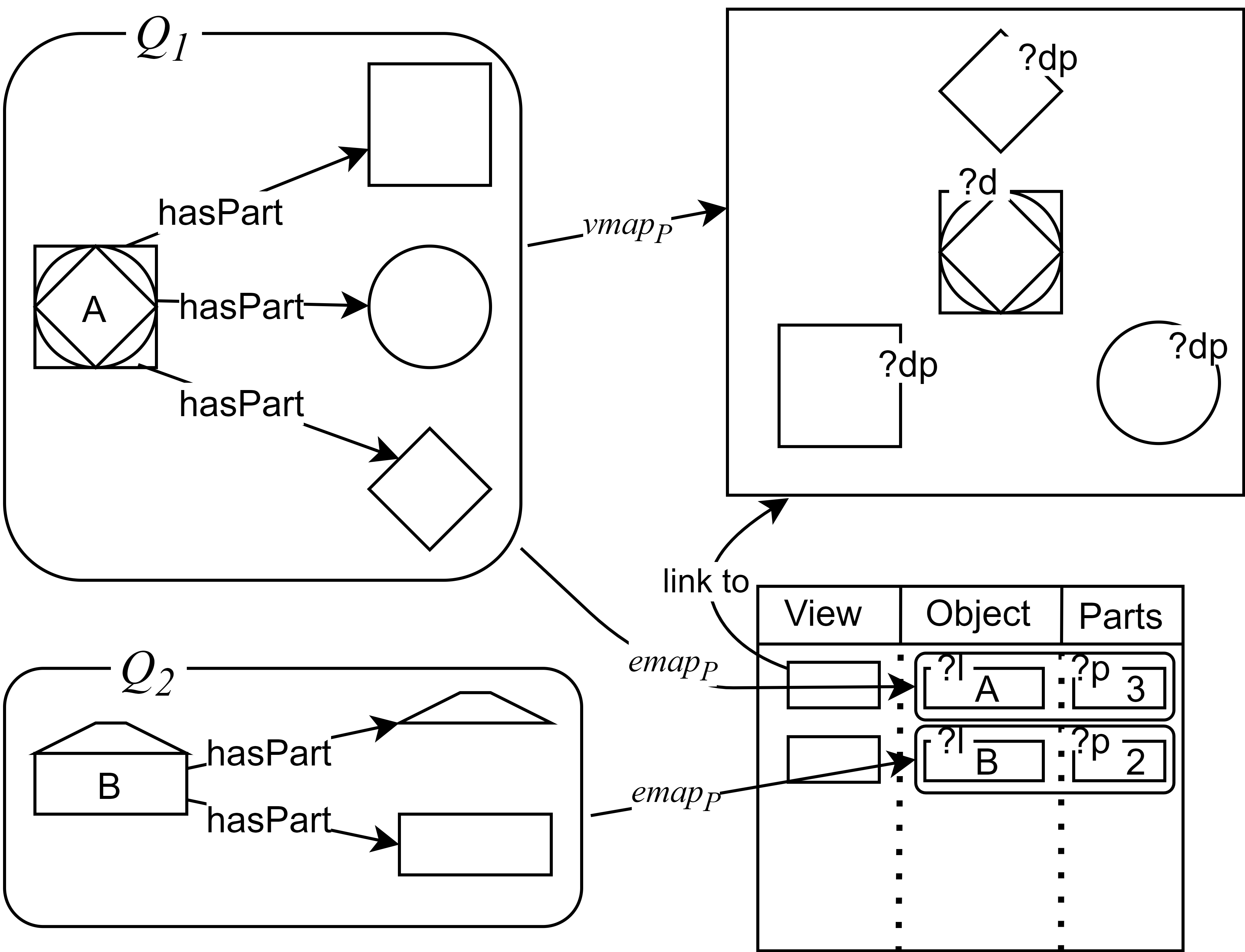}
         \caption{Example of pattern instances mapping}
         \label{fig:partof_mapping}
     \end{subfigure}
        \caption{An example of application of our approach on pattern Part Of}
        \label{fig:partof_example}
\end{figure}

Note that our approach is completely independent of the specific implementation of patterns in the knowledge graph and therefore enable reusing the visualization code of a pattern on a different knowledge graph that implements the same pattern and is annotated with the extended OPLa. To make a pattern $P$ visualizable, the developer needs to compose the set of filter controls, design the visual frame and define the functions $emap_P(Q)$ and $vmap_P(Q)$. Figure \ref{fig:partof_example} shows an example of a pattern and how data of specific instances is mapped into the GUI. The left part (Figure \ref{fig:partof_pattern}) depicts the \emph{Part Of} ODP, defined by property \texttt{hasPart}, which connects an object to its parts. We include the properties \texttt{label} and \texttt{depiction}, although not part of the pattern, since they  are present (at least label) in all KGs for crucially increase their human-readability. The pattern is associated to the $emap_P$ and $vmap_P$ functions, whose implementation is reported in the right side of Figure \ref{fig:partof_pattern} in pseudo-SPARQL. Given a pattern instance $Q$, identified by uri($Q$), $emap_P(Q)$ retrieves the object label (\texttt{?l}) and the number of parts (\texttt{?p}). It also applies the filter (\texttt{user\_defined\_filter} in the WHERE clause) and return $\phi$ (an empty set of tuples) if filter constraints are not satisfied. $vmap_P(Q)$ extracts figures of the object and every part (\texttt{?d} and \texttt{?dp}, identified by property \texttt{depiction}) and maps them into the visual frame (the visual mapping is summarized by the $visual\_frame()$ function). The right part (Figure \ref{fig:partof_mapping} shows a concrete example of two instances $Q_1$ and $Q_2$ of pattern Part Of and their mapping into the table for the exploration (bottom-right side) and the visual frame (top-right side). For space reasons we only show the visual frame for $Q_1$. For each label and depiction we report the corresponding SPARQL variable of the $emap_P$ and $vmap_P$ implementations (taken from Figure \ref{fig:partof_pattern}).  

\hide{

\subsection{Visualization of specific ODPs}

The proposed framework can be extended with new ODPs by implementing the pattern-specific parts, i.e. composition of filter controls, implementation of the visualization frame, definition of the functions $emap_P(Q)$ and $vmap_P(Q)$. We describe the   

\bigbreak

import) c coincide c
specialization) c1 specializza c if c1 imports c and at least one ontology element of c1 specialize one of c
generalization) c1 generalizza c if c1 imports c and at least one ontology element of c1 generalize one of c
composition) association classes/properties of c1 with c2 

specializzazione del pattern Collection 

?x memberOf ?y
?y subClassOf Collection

?e a Entity
?c a Collection
?e isMemberOf Collection

?e a ?z
?c a ?t
?e ?p ?c
?t subClassOf* Collection 
?z subClassOf* Entity
?p subClassOf* isMemberOf

capire quando è necessaria clausola GROUP BY

G = (V U P, E, ln, le) 
}

%% file: src/implementation.tex
\noindent
\textit{Design Methodology.}
We put emphasys on the importance of user cognition in the visualization design process~\cite{patterson2014human} and, taking inspiration by other successful  experiences (e.g.~\cite{he2019aloha})
we applied User-Centered Design (UCD) principles. 
Specifically, we followed agile design methodology  with brief iteration cycles (1 week sprints). 
This process involved 2 developers, and 3 users collaborated in the process by continuously providing feedback on the tool.
Once the tool reached a stable version, we conducted an experiment with an 11-participants focus group and evaluated the usability according to the System Usability Scale (SUS) \cite{brooke2013sus}.
The results  are discussed in Section \ref{sec:evaluation}.

\noindent
\textit{ODPReactor Architecture.}
\emph{ODPReactor}  is divided into three main components: 
\begin{itemize}
    \item \emph{ODPBrowser}, a service to explore and filter the ODP/key concepts graph and their instances (it encapsulates ODP level and exploration level described in Sect. \ref{sec:approach}); 
    \item \emph{Extended-Ld-Reactor (E-LD-R)}, a service to visualize resource pages with their data (it encapsulates the visualization level); 
    \item \emph{ODP-UI}, the visual frames library, embedded in the \emph{E-LD-R} microservice.
\end{itemize}

\begin{figure}[t]
\centering
  \includegraphics[width=0.7\linewidth]{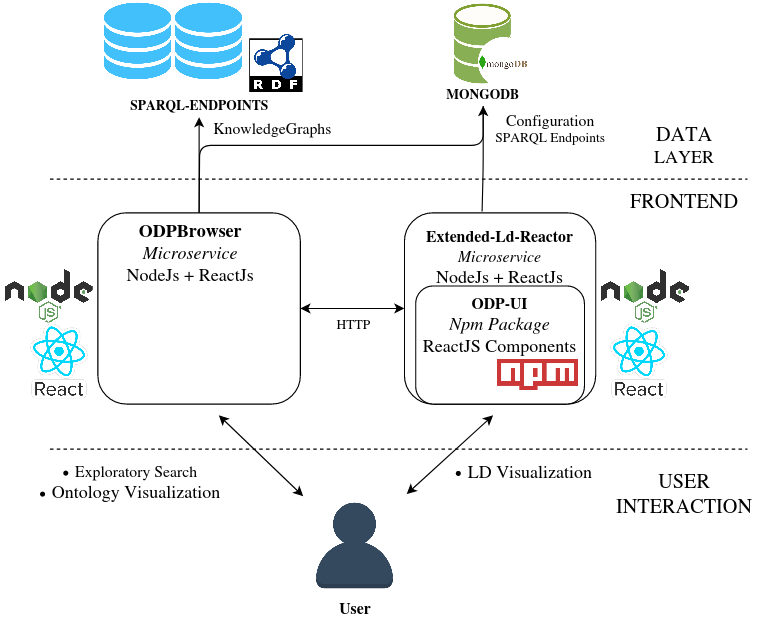}
  \caption{ODPReactor architecture.}
  \label{fig:architecture}
\end{figure}

The two services are exposed in a client-side React.js\footnote{\url{https://reactjs.org/}} application hosted and served by a Node.js\footnote{\url{https://nodejs.org/}} environment. The services can be configured to retrieve data from one or more SPARQL endpoints and store configuration data (e.g. the list of KGs and their respective SPARQL endpoints) in a MongoDB\footnote{\url{https://www.mongodb.com/}} 
The architecture is detailed in Figure \ref{fig:architecture}.
We discuss here the three components.

\textbf{ODPBrowser} is a service implementing the ODP level (ODPs and concepts graph visualization) and the exploration level (table with filters) discussed in Sect. \ref{sec:approach}.
The ODP/key concepts graph is implemented by using Graphin\footnote{\url{https://graphin.antv.vision/}}, a graph visualization library. 
Every concept has an associated importance score (computed by \emph{degree centrality}), which enables filtering concepts whose importance is below a user-defined threshold. 
By double clicking on a node, the user can navigate to the exploration level 
where all instances of the selected key concept or data associated to the selected ODP are displayed in a table. 
This view contains filters semantically related to the specific ODP (e.g. the ODP \emph{Time Indexed Situation} enable filtering by time interval). 

\textbf{ODP-based Semantic Filtering.} To filter geographic locations, we implemented a map with LeafletJS\footnote{\url{https://leafletjs.com/}} where users can draw a closed perimeter. 
Perimeter coordinates are then calculated and resources localized inside the area are selected while others discarded. Numeric filtering (such as measures, time intervals, number of components) can be performed by means of sliders where the user can select with a cursor interval of interest. We used React Compound Slider\footnote{\url{https://react-compound-slider.netlify.app/}} to develop the facets. Among other facets, we used checkboxes to select or deselect categorical data such as location types.
By double clicking on a table row, the user is redirected to \emph{Extended-Ld-Reactor}, which visualizes the selected instance.

\textbf{Extended-Ld-Reactor} is an extension of Linked Data Reactor (LD-R)~\cite{khalili2016linked}, a full stack application (based on Fluxible\footnote{\url{https://fluxible.io/}}, React.js and Semantic UI\footnote{\url{https://semantic-ui.com/}}), which facilitates mapping URIs into visual components. 
We embedded specific user interface for ODP-based visualization  in LD-Reactor and configured it to visualize resources related to ODPs. The user interface makes use of a library of visual frames (namely \emph{ODP-UI}, discussed below), which can be visualized singularly or through a mosaic grid with all visual frames associated to a specific resource.

\textbf{ODP-UI} is a library of visual frames where each frame displays data related to a specific ODP, as discussed in Sect. \ref{sec:approach}. Each visual frame is implemented as a React.js component. We implemented visual frames for the patterns Time-Indexed Typed Location, Measurement Collection, and Cultural Property Component Of, discussed in Sect. \ref{sec:background}.

\noindent
\textit{Deployment.}
We deployed ODPReactor for the ArCo knowledge graph \cite{carriero2019arco}. ArCo is the Italian Cultural Heritage knowledge graph, consisting of a network of seven vocabularies and 169 million triples representing 820 thousand cultural entities. 
The ArCo ontology has been developed following the eXtreme Design methodology, an ontolology design methodology which uses ODPs as building blocks for constructing ontologies.
This makes ArCo a good candidate for evaluating our approach.

\noindent
\textit{Code.} The  implementation of the framework is available as open source project on GitHub\footnote{\url{https://github.com/ODPReactor}}.


\noindent
\textit{Use case scenario.}
We describe a use-case scenario as an example to show how the  \emph{ODPReactor} components interact. 
A user (say Claire) is interested in getting information about all cultural properties that in 1856 were located in a specific area in Paris. 
Claire accesses to the \emph{ODPBroswer}, retrieves the ODPs composing ArCo and double-clicks the node representing the Time-Indexed Typed Location. 
A data table with cultural property instances and associated data (according to dimensions of the pattern) is shown and a series of filter facilities are made available to her. 
Claire then filters data by: (i) drawing an area around Paris on the location filter control geographical map; and (ii) selecting a time interval with the interval filter control slider. She locates the specific cultural property she is interested in, and selects it. 
\emph{ODPBrowser} pass the control to \emph{Extended-Ld-Reactor} to load the instance data from the SPARQL endpoint and display it in a mosaic grid of visual frames. 

\reminder{quest'esempio mi sembra non aggiungere niente alla discussione, forse andrebbe rimosso}

\hide{
\begin{figure}[h!]
  \includegraphics[width=\linewidth]{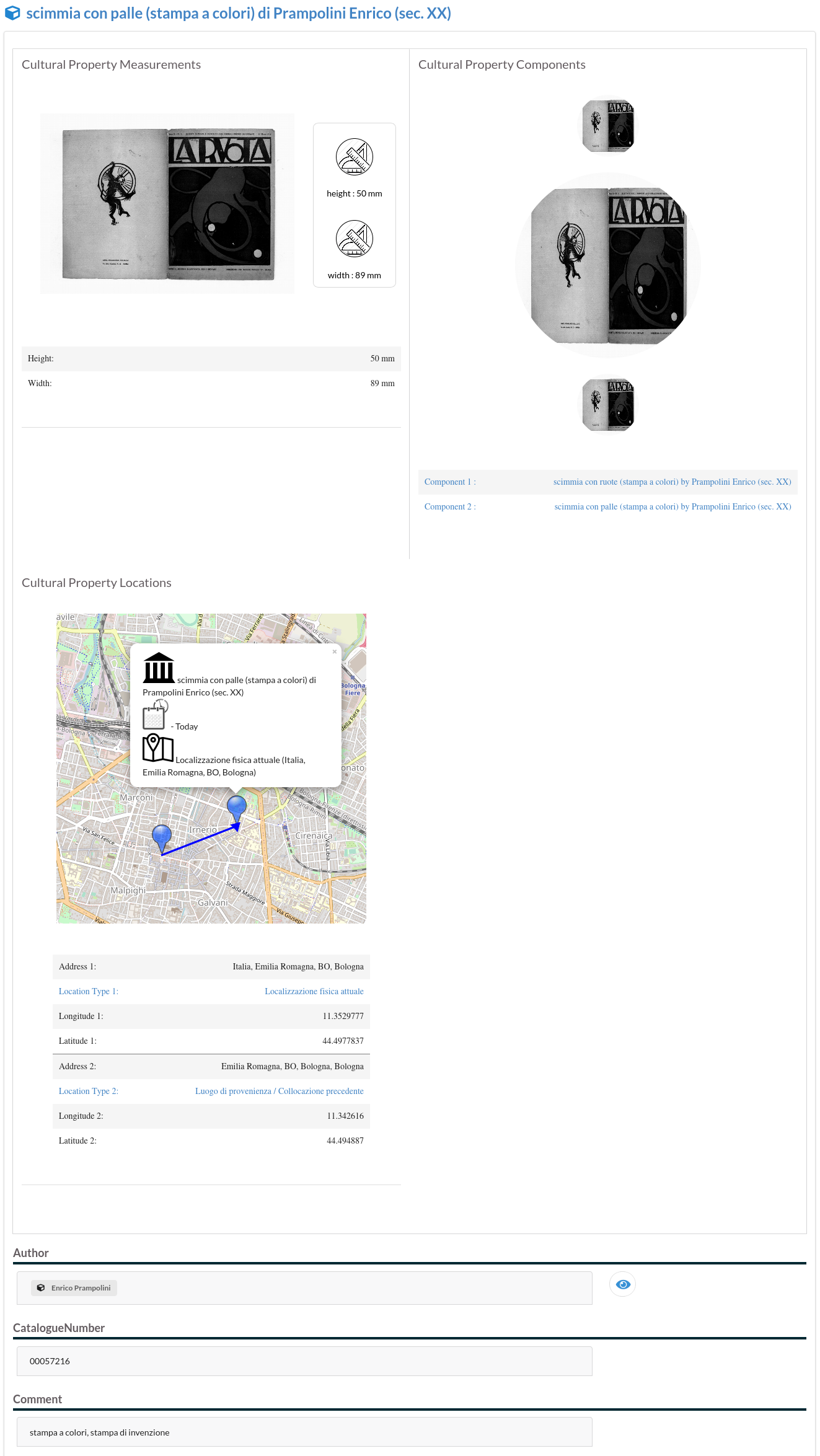} 
  \caption{ODP-Reactor: visual frame mosaics to display resource data.}
  \label{fig:mosaic-vie}
\end{figure}
}

%% file: src/evaluation.tex
In this section, we report on the results of the user evaluation of the proposed approach. 
As proof-of-concept of our approach, we deployed ODPReactor for a portion of the ArCo KG.
The portion included 200  cultural properties randomly selected among those participating to (at least) an instance of the three ODPs chosen for the case study.
In terms of pattern instances this selection included: 49 instances for \textit{Cultural Property Component Of}, 270 for \textit{Time Indexed Typed Location} and 158 for  \textit{Measurement Collection}. 
The test was conducted by 11 users, composed by 3 researchers, 5 Phd students and 3 working professionals. 
All participants were familiar with Semantic Web and Linked Data technologies and were not involved in the development process.

The purpose of the test was to evaluate the ability of the tool to perform the following tasks:
\begin{itemize}
    \item present the KG in a clear, concise and intuitive way (cf. \textbf{H1}, \textbf{H2}, \textbf{H4});
    \item give the users a usable interface for searching  specific instances (cf. \textbf{H3}).
\end{itemize}

Each tester was asked to perform a series of searching tasks (e.g. finding the length of an object) and to answer questions about the selected items.
We dedicated a section of the questionnaire for each task.
Each section presents the task and asks the user to rate:
\begin{enumerate*}[label=\textit{(\roman*)}]
    \item the difficulty and rapidity in performing the task;
    \item the usability of the tool.
\end{enumerate*}
The scores range was on a scale of 1 (very easy/rapid/usable) to 5 (very difficult/slow/unusable).
To evaluate the general level of user satisfaction, we added to the survey a section containing the SUS-System Usability Scale questionnaire \cite{brooke1996}.
Before starting the test, the users were asked to perform a brief tutorial defined for the tool. 
The tutorial presents the user the three visualization levels (ODP level, exploration level, visualization level) and, in each of them, it shows the position of the filters.
We measured the actual time occurred for performing each task and asked the participants to record their screen and send us the recording. The latter was useful for understanding the reason of possible failures or slowdowns. 
10 of the 11 participants agreed to record their screen and share the recording.

\begin{figure}[t!]
\centering
  \includegraphics[width=0.4\linewidth]{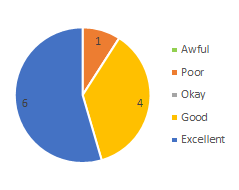}
  \caption{SUS results.}
  \label{fig:sus}
\end{figure}

The results of the evaluation according to the SUS scale are presented in Figure \ref{fig:sus}. They show that the tool is considered excellent by more than half of the users and good by the remaining ones except one, whose judgment was poor. The average score was $81.4$, with a standard deviation of $12.6$.

\noindent
\textit{Assessment of the visualization level.} A first group of tasks, reported in  Table \ref{tab:visualization_tasks}, was designed to evaluate the clarity of the visualization level (therefore assessing the hypothesis \textbf{H4}). 
In order to compare ODPReactor with existing tools we asked each user to perform the same task also using LodView (a widely known KGs viewer adopted for ArCo).
We considered one task per ODP View. 
Each task was executed twice per tool in order to measure also the improvement with the experience. For each execution the user was asked questions about the dimensions of the ODP.

\begin{figure}[t!]
  \centering
  \begin{subfigure}[b]{0.40\linewidth}
    \includegraphics[width=\linewidth]{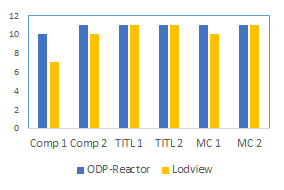}
     \caption{Correctness}\label{fig:visual_correctness}
  \end{subfigure}
  ~
  \begin{subfigure}[b]{0.40\linewidth}
    \includegraphics[width=\linewidth]{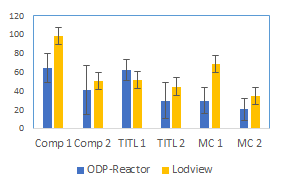}
    \caption{Timing.}\label{fig:visual_timing}
  \end{subfigure}
  ~
  \begin{subfigure}[b]{0.50\linewidth}
    \includegraphics[width=\linewidth]{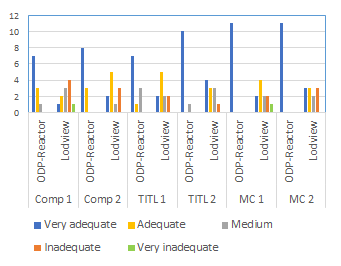}
    \caption{Usability.}\label{fig:visual_usability}
  \end{subfigure}
  \caption{Data visualization tasks}
  \label{fig:data_visual}
\end{figure}

We considered as evaluation metric the number of correct answers, the time of execution and the users judgment of usability. 
An answer has been evaluated as correct if the user correctly identified the entry containing the data, although the answer is only partial. 
For example in tasks TITL1 and TITL2, for little-known cities the users had difficulties in distinguishing the city from the village or from the county (all reported in the same entry). \todo{Quanti sono?}
Therefore we considered all such answers as correct. The results are reported in Figure \ref{fig:data_visual}.
Figure \ref{fig:visual_correctness} shows the number of correct task executions. ODPReactor always performs similarly or better than LodView. 
Figure \ref{fig:visual_timing} shows the average task execution time. ODPReactor always performs better than LodView, with the exception of TITL 1, which regards retrieving the location of a cultural property. 
By analyzing the recordings it emerges that the users have quickly identified the data, but have delayed the transcription, because of difficulties in distinguishing the town (specifically Caravino and Vittorio Veneto) from the village or from the province. The same behaviour is not reflected in the LodView tasks, since the answers consist in widely-known cities (specifically Rome and Turin).
Figure \ref{fig:visual_usability} shows the user judgment of the tool, where ODPReactor was considered significantly more adequate than LodView. This result is emphasized for the Measurement Collection tasks, where ODPReactor was evaluated unanimuosly very adequate while the judgment for LodView was almost equally distributed from Very adequate to Inadequate.


\begin{table}[t!]
\centering
\begin{tabular}{|l|c|c|}
\hline
ID     & ODP involved & Question \\ 
\hline
Comp 1 & \multirow{2}{*}{\begin{tabular}[c]{@{}c@{}}Cultural Property \\ Component of\end{tabular}} & \multirow{2}{*}{\begin{tabular}[c]{@{}c@{}}How many components is the object \\ made of?\end{tabular}}          \\ \cline{1-1}
Comp 2 & &  \\ \hline
TITL 1 & \multirow{2}{*}{\begin{tabular}[c]{@{}c@{}}Time   indexed typed \\ location\end{tabular}} & \multirow{2}{*}{\begin{tabular}[c]{@{}c@{}}In which city is the cultural property \\ located?\end{tabular}}     \\ \cline{1-1}
TITL 2 & & \\ \hline
MC 1   & \multirow{2}{*}{Measurement collection} & \multirow{2}{*}{\begin{tabular}[c]{@{}c@{}}What is the length/height of this cultural \\ property?\end{tabular}} \\ \cline{1-1}
MC 2   &                                                                                           &                                                                                                                 \\ \hline
\end{tabular}
\caption{Data visualization tasks}\label{tab:visualization_tasks}
\end{table}

\noindent
\textit{Assessment of the ODP and exploration level.}
In order to assess  \textbf{H1-3} hypotheses we designed a  group of tasks involving  the ODP level and exploration level. These tasks, reported in Table \ref{tab:exploration_tasks}, are designed for evaluating the benefits of the approach in streamline the navigation and search of information in the KG.
All the tasks required to identify the number of occurrences in the KG satisfying a given searching criterion.
The first two tasks (referred to \textit{Measurement collection} ODP and \textit{Cultural property component of} ODP) can be performed by applying just one filter, while the last three ones (referred to \textit{Time indexed typed location} ODP) require the activation of more than one filter. We chose the ODP Time Indexed Typed Location for the multi-filter tasks since this pattern is more complex then the others and hence it has more filter controls.
To evaluate the general level of user satisfaction, we asked the users to answer a survey  defined according to the SUS-System Usability Scale questionnaire~\cite{brooke1996}.

\begin{table}[t!]
\centering
\begin{tabular}{|c|c|c|}
\hline
ID        & ODP involved                                                                           & Question                                                                                                        \\ \hline
ES Task 1 & Measurement collection                                                                 & \begin{tabular}[c]{@{}c@{}}How many cultural properties are higher \\ than 2 m?\end{tabular}                    \\ \hline
ES Task 2 & {\begin{tabular}[c]{@{}c@{}}Cultural Property \\ Component of\end{tabular}}                                                                             & \begin{tabular}[c]{@{}c@{}}How many cultural   properties have at least \\ eight components?\end{tabular}       \\ \hline
ES Task 3 & \multirow{2}{*}{\begin{tabular}[c]{@{}c@{}}Time indexed typed\\ location\end{tabular}} & \begin{tabular}[c]{@{}c@{}}How many cultural properties have Firenze as \\ their current location?\end{tabular} \\ \cline{1-1} \cline{3-3} 
ES Task 4 &                                                                                        & \begin{tabular}[c]{@{}c@{}}How many cultural properties were there in \\ Firenze before the 1945?\end{tabular}  \\ \cline{1-1} \cline{3-3} 
ES Task 5 &                                                                                        & \begin{tabular}[c]{@{}c@{}}How many works by Prampolini Enrico are \\ there in Bologna?\end{tabular}            \\ \hline
\end{tabular}
\caption{Exploratory search tasks}\label{tab:exploration_tasks}
\end{table}


Finally, we report on the results of the exploratory research tasks on ODPReactor. We did not compare this task with LodView since it does not provide exploratory search functionalities. In counting the number of occurrences that satisfy the given constraints, we did not make any assumption about open or close world, and considered correct both the answers that include or exclude objects that do not contain some of the properties necessary for applying the filter.

\begin{figure}[t!]
  \centering
  \begin{subfigure}[b]{0.30\linewidth}
    \includegraphics[width=\linewidth]{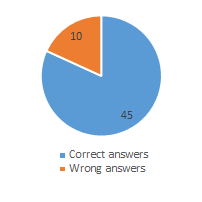}
     \caption{Correctness.}\label{fig:explor_correctness}
  \end{subfigure}
  ~
  \begin{subfigure}[b]{0.32\linewidth}
    \includegraphics[width=\linewidth]{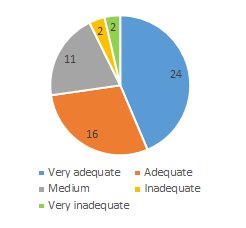}
    \caption{Usability.}\label{fig:explor_usability}
  \end{subfigure}
  ~
  \begin{subfigure}[b]{0.32\linewidth}
    \includegraphics[width=\linewidth]{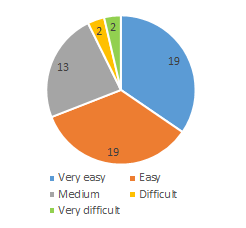}
    \caption{Task difficulty.}\label{fig:explor_difficulty}
  \end{subfigure}
  \caption{Exploratory search tasks}
  \label{fig:exploration}
\end{figure}

Results are summed up in Figure \ref{fig:exploration}. We collected $55$ ($11 \times 5$) responses by aggregating correctness, usability and difficulty, from all tasks and all users. Most of the users performed the task correctly (Fig. \ref{fig:explor_correctness}). We analyzed the recordings for understanding the reasons of failure or difficulties in executing the task. It emerged that among 10 wrong answers, 9 ones are due to failures to apply all filters necessary for the complex multi-filter tasks, while for the other one the data was correctly identified but wrongly reported. 
Specifically, the testers wrongly set the filter of the Time-indexed Typed Location instances for selecting only those having a certain type (e.g. Current Location).
This indicates that the usage of such filter was not intuitive for the testers and it has to redesigned.
Finally, Figure \ref{fig:explor_usability} shows the result of the questions about usability of the tool. 
Most of the users judged the tool very adequate or adequate. 51 answers out of 55 were positive or intermediate.
Figure \ref{fig:explor_difficulty} shows similar positive results about difficulty. 


%% file: src/conclusions.tex
This work proposed a novel approach to Knowledge Graph visualization that uses Ontology Design Patterns as first-class citizens for accessing and navigating KGs.
We described a general framework that enables reusing an ODP-related visual frame to different KGs and implemented a tool for validating the concept. 
The results of the execution of a set of tasks on the tool demonstrated the validity of the proposed approach, confirmed by considering both objective parameters and user perspective.
In future we plan to enrich our library of visual frames with new ODP-based views and to explore general methods for automatizing the annotation of pattern instances. 
We also plan to study an automatic system for building  filter facets and to make an extensive evaluation on a larger number of users.